\RequirePackage{amsmath}
\documentclass[runningheads]{llncs}
\usepackage{graphicx}
\usepackage{makeidx}
\usepackage{amssymb}
\usepackage{mathtools}
\usepackage{amsfonts}
\usepackage{dsfont}

\usepackage[normalem]{ ulem }
\usepackage{soul}

\usepackage{hyperref}

\usepackage{subcaption}

\usepackage{stmaryrd}
\usepackage{hhline}

\usepackage[table]{xcolor}

\newcommand{\bx}{\mathbf{x}}

\newcommand{\R}{\mathbb{R}}
\newcommand{\Z}{\mathbb{Z}}

\begin{document}

\title{Scale Equivariant Neural Networks with Morphological Scale-Spaces}

\titlerunning{Scale Equivariant Neural Networks with Morphological Scale-Spaces}
\authorrunning{M. Sangalli \and S. Blusseau \and S.Velasco-Forero \and J. Angulo}
\author{Mateus Sangalli \and Samy Blusseau \and Santiago Velasco-Forero \and Jesús Angulo}
\institute{Centre for Mathematical Morphology, Mines Paristech, PSL Research University, France}
\maketitle              
\begin{abstract}

 The translation equivariance of convolutions can make convolutional neural networks translation equivariant or invariant.
 Equivariance to other transformations (e.g. rotations,  affine transformations, scalings) may also be desirable as soon as we know \emph{a priori} that transformed versions of the same objects appear in the data.
The semigroup cross-correlation, which is a linear operator equivariant to semigroup actions, was recently proposed and applied in conjunction with the Gaussian scale-space to create architectures which are equivariant to discrete scalings.
In this paper, a generalization using a broad class of liftings, including morphological scale-spaces, is proposed.
The architectures obtained from different scale-spaces are tested and compared in supervised classification and semantic segmentation tasks where objects in test images appear at different scales compared to training images.
In both classification and segmentation tasks, the scale-equivariant architectures improve dramatically the generalization to unseen scales compared to a convolutional baseline.
Besides, in our experiments morphological scale-spaces outperformed the Gaussian scale-space in geometrical tasks.

\end{abstract}

\section{Introduction}

\begin{figure}
    \centering
    \begin{subfigure}{.49\textwidth}
        \centering
        \begin{tabular}{cccc}
            \includegraphics[width=.25\textwidth]{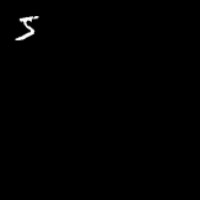} &
            \includegraphics[width=.25\textwidth]{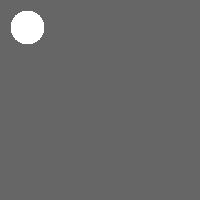} &
            \includegraphics[width=.25\textwidth]{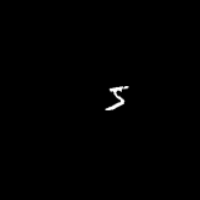} &
            \includegraphics[width=.25\textwidth]{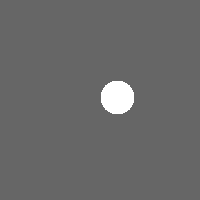}
        \end{tabular}
        \caption{}
    \end{subfigure}%
    ~
    \begin{subfigure}{.49\textwidth}
        \centering
        \begin{tabular}{ccc}
            \includegraphics[width=.25\textwidth]{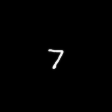} &
            \includegraphics[width=.25\textwidth]{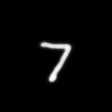} &
            \includegraphics[width=.25\textwidth]{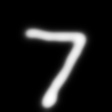}
        \end{tabular}
        \caption{}
    \end{subfigure}%
    \caption{Image from MNIST \cite{mnist} and their translated versions with a illustration of their respective feature maps in a CNN(a), and images from MNIST Large Scale \cite{jansson2020unseenscales} at different scales(b).}
    \label{fig:example_mnist_images}
\end{figure}

Convolutional Neural Network (CNN) models achieve state-of-the-art performance in many image analysis tasks. An important property of CNNs is that a translation applied to its inputs is equivalent to a translation applied to its features maps, as illustrated in Fig. \ref{fig:example_mnist_images}. This property is a particular case of group equivariance \cite{cohen2016group}. An operator is equivariant with respect to a group if applying a group action in the input and then the operator, amounts to applying the operator to the original input and then an action of the same group to the outputs. In addition to translations, group actions can model many interesting classes of spatial transformations such as rotations,  scalings, affine transformations, and so on.

Group equivariant CNNs~\cite{cohen2016group} are a generalization of CNNs that, in addition to being equivariant to translations, are also equivariant to other groups of transformations. Many of these networks focus on equivariance to rotations, in different kinds of data \cite{cohen2016group,weiler2018steerable,weiler20183d,thomas2018tensor}. 
A group equivariant neural network may also be used to obtain invariance, with reduction operations. An operator is invariant to some transformation if applying the operator to an input or to its transformed version produces the same output. Invariance is often crucial in image analysis tasks. For example, in a digit classification task a translation should not change the label of the digit, as illustrated by Fig. \ref{fig:example_mnist_images}(a). The same holds for re-scaled versions of the same digit (Fig. \ref{fig:example_mnist_images}(b)).

Worrall and Welling~\cite{worrall2019scale} introduce neural networks equivariant to the action of semigroups, instead of groups. Semigroup actions can model non-invertible transformations and in~\cite{worrall2019scale} the authors focused on equivariance to downsampling in discrete domains. Focusing on downsampling is a way to address equivariance to scalings without creating spurious information through interpolation. In their architecture, the first layer is based on a Gaussian scale-space operator, and subsequent layers of this network are equivariant to the action of a semigroup of scalings and translations. Effectively, these operators are equivariant to rescaling of a discrete image. There are other scale-spaces with similar mathematical properties to the Gaussian scale-space, in particular the morphological scale-spaces.

In this paper we generalize the approach on scale-equivariant neural networks~\cite{worrall2019scale} by finding a sufficient condition in which scale-spaces, in the sense of~\cite{heijmans02scale}, can be used as the first layer, or the so-called \emph{lifting}, in an equivariant network, and investigate several architectures built on morphological scale-spaces.
We observe that the morphological scale-spaces networks compare favorably to the Gaussian one in tasks of classification and segmentation of images at scales previously unseen by the network, in contrast to \cite{worrall2019scale}, in which the experiments test the overall performance of the network, but where the train and test sets objects follow the same scale distribution.
The rest of the paper is organized as follows.
In Section~\ref{sec:related-work} we give a short review of the existing approaches related to equivariance in CNNs. Then the general mathematical framework we use is exposed in Section~\ref{sec:general-setting}, before we focus specifically on the semigroup of scalings and translations in Section~\ref{sec:scale-translation-equivariance}. In the latter, we also review the algebraic basis of scale-spaces \cite{heijmans02scale} and apply this definition to generalize the scale-equivariant architecture of \cite{worrall2019scale}.
In Section~\ref{sec:experiments} we test scale-equivariant architectures obtained from different scale-spaces. In particular we test the models in classification and segmentation of images where the objects in the test set appear at scales unseen in the training set. We opt to test the models in tasks where the scale of objects can be easily controlled and measured rather than tasks with real data, where even though objects may appear at different scales, it is difficult to explicitly differentiate between the training and test set scales'. We focus on simple experiments which depend on the shapes of the objects, rather than textures. In those experiments, the equivariant models improved the generalization accuracy dramatically over the convolutional models and the morphological scale-spaces' models performed well even in comparison to the Gaussian scale-spaces.

\section{Related Work}
\label{sec:related-work}

In \cite{zhu2019scale}, scale equivariance is obtained by applying filters in the different scales using decomposed kernels to reduce the model complexity. The authors applied the equivariant model to multiscale classification and reconstruction of hand-written digit images and were able to surpass the regular CNN models in the generalization to unseen scales, even when using data augmentation. 
In \cite{ghosh2019scale} a locally scale invariant neural network architecture is defined. Filters are defined as linear combinations of a  basis of steerable filters and max-pooling the result over different scales. The invariant network was successfully applied to the tasks of classification of re-scaled and distorted images of hand-written digits and clothing. Both of these approaches reduce computational cost and avoid creating spurious information through interpolation by using a decomposition into steerable filters, but applying these models to large scale variations can increase their cost significantly.
In \cite{jansson2020unseenscales}, a scale invariant architecture is proposed, in which input images are processed in different scales, with foveated images operators, i.e. images are processed with a higher resolution close to the center and a smaller resolution as the operators gets farther from the center.
The MNIST Large Scale dataset, which is used later in Section \ref{sec:experiments} was introduced there, and the foveated networks achieved very high generalization performance in unseen scales when compared to regular CNNs. A disadvantage of this approach is that it assumes that the objects of interest are at the center of the image.

In \cite{worrall2019scale}, instead of treating scaling as an invertible operation, such as it would behave in a continuous domain, it is considered the action of downsampling the input image in a discrete domain. Because of that, the obtained operators are equivariant to a \emph{semigroup}, and not a group. The semigroup action in the input consists of applying a scale-dependent Gaussian blur to the inputs and then downsampling, which is the way to re-scale discrete signals while avoiding aliasing. The convolutional filters are efficiently applied to feature maps defined on a semigroup formed by scalings and translations by means of dilated convolutions, without relying on interpolation. These operators are also easily scalable to large scale variations, since applying it at larger scales has the same computational cost. The semigroup equivariant models were applied to classification and semantic segmentation of datasets of large images, achieving results which are competitive with the literature.
The Gaussian scale-space may not be appropriate when blurring affects the geometrical features that characterize the objects to analyse.
With this in mind, the models in this paper are an extension of the ones in \cite{worrall2019scale}, but we generalize the approach to allow the usage of other scale-spaces \cite{heijmans02scale}, with a focus on morphological scale-spaces, as a step to the generalization of this model to more complicated data, as well as a way to shed some light into the workings of morphological scale-spaces in the context of these images.

\section{General setting}
\label{sec:general-setting}

In the scope of image processing, the notion of equivariance of
an operator means that a transformed version of an image should
produce an ``equivalently'' transformed version of the original output
by the operator, as illustrated in
Fig.~\ref{fig:example_mnist_images}(a). We are specifically
interested in \emph{linear} operators, as they are the elementary
operations in common neural networks.

\subsection{Group equivariance}
\label{sec:group-equivariance}

Let $(G, .)$ be a discrete group and $\mathcal{F} = \mathbb{R}^G$ the set
of functions mapping $G$ to $\mathbb{R}$. Consider the family of
operators $R_g, g\in G$ defined on $\mathcal{F}$ by
\begin{equation}
    \forall g\in G, \forall f\in \mathcal{F}, \;\;\; R_g(f) : u\in G \mapsto f(u.g^{-1}).
\end{equation}
This family of operators is a right group action of $G$ on
$\mathcal{F}$ (as $R_{g_1}\circ R_{g_2} = R_{g_2 g_1}$).  For
illustration, the group $G$ could be for example the group of
translations of $\mathbb{Z}^2$, identified to $(\mathbb{Z}^2, +)$
itself, and in that case $\mathcal{F}$ could be seen as the set of
``infinite" images (or classical images periodised over all
$\mathbb{Z}^2$). In turn, the action $R_g$ would be the translation of
a function by a vector $g$.

Bearing in mind the final purpose of defining CNN layers, we focus on linear endomorphism of the vector space $\mathcal{F}$. Suppose such an operator $H$ is equivariant with respect to $R_g$, that is,
\begin{equation}
    \forall f\in \mathcal{F}, \;\;\; H \left(R_g(f)\right) = R_g \left(H(f)\right).
\end{equation}
Then, using linearity and the fact that the basis
$(\mathds{1}_{\lbrace g \rbrace})_{g\in G}$ spans $\mathcal{F}$, we get
$H(f) = \sum_{g\in G} f(g) R_g (h)$
where $h = H(\mathds{1}_{\lbrace e_G \rbrace})$ is the response of the
filter $H$ to the impulse on $e_G$, the neutral element of $G$. This
writes in the more familiar form
\begin{equation}
  \label{eq:group-convolution}
  \forall u\in G, \;\;\;  H(f)(u) = \sum_{g\in G} f(g) h(u.g^{-1}) = \sum_{g\in G} f(g^{-1}.u) h(g).
\end{equation}
We end with a classical result, at the basis of linear filtering and
convolutional neural networks: linearity and equivariance implies for an
operator to be written as a convolution with a kernel $h$ that
represents it (and conversely). We shall note
$H(f):= f \star_G h $ although this operation is commutative only if $G$
is.

\subsection{Semigroup equivariance}
\label{sec:semi-group-equivariance}

Let's first stress the interest of extending the equivariance
setting to semigroups. Recall that $(G, \cdot)$ is a semigroup if the law $\cdot$ is associative, but in general they do not have a neutral element or inverse elements. A semigroup induces a
semigroup action $(\varphi_g)_{g\in G}$ on a set $X$, as soon as this
family is homomorphic to the semigroup, that is, if either
$\forall g, h \in G$,
$\varphi_g \circ \varphi_h = \varphi_{g\cdot h}$.
or
$\forall g, h \in G$,
$\varphi_g \circ \varphi_h = \varphi_{h\cdot g}$.

In image processing, important examples of semigroup actions are
scale-spaces. As we will present in more details in
Section~\ref{sec:scale-spaces}, semigroup actions
on images may be the convolution with a Gaussian kernel (Gaussian scale-space) or the application of a morphological operator such as erosion, dilation, opening or closing (morphological scale-spaces). Scale-spaces highlight the multi-scale nature of images and have shown great efficiency as image representations
\cite{lowe1999object}. Besides, they are
naturally complementary to the scaling operation. For example, the Gaussian blurring acts as a low-pass filter and allows the subsampling (or downscaling on a discrete domain) of an image to avoid aliasing artifacts.

Hence, equivariance of linear operators to semigroups seems highly
desirable, as it is natural to expect that the same information at
different scales produce the same responses up to some shift due to
scale change. However, the derivation of
Section~\ref{sec:group-equivariance} cannot be reproduced here since
it includes group inversions, which precisely lack in
semigroups. Still, we notice that
Equation~\ref{eq:group-convolution} can also be written
$H(f)(u) = \sum_{g\in G} f(u.g) h(g)$
if we change the function $h$ for its conjugate, that is
$h(u) = H(\mathds{1}_{\lbrace e_G \rbrace})(u^{-1})$, and thanks to a
change in variables. Now this expression can be applied to semigroups,
considering the semigroup right action $R_u(f)(g) = f(u.g)$. We get
that operators $H$ defined by
\begin{equation}
  \label{eq:semigroup-convolution}
  \forall u\in G, \;\;\;  H(f)(u) = \sum_{g\in G} R_u(f)(g) h(g)
\end{equation}
are indeed equivariant to the semigroup action $R_t, t\in G$, since
\begin{equation}
    H(R_t(f)) (u) = \sum_{g\in G} R_u(R_t(f))(g) h(g) = \sum_{g\in G}
R_{tu}(f)(g) h(g) = R_t(H(f))(u).
\end{equation}
This class of semigroup equivariant operators is the semigroup cross-correlation proposed in~\cite{worrall2019scale}. We also write $f \star_G h$, remarking however that, contrary to the group case, this operation is not
symmetrical in $f$ and $h$ even when the law $\cdot$ on $G$ is
commutative.

\subsection{Lifting}
\label{sec:lifting}

So far we have considered functions defined on a general semi-group,
but the input to CNNs are images defined on a discrete set $X$, which
may be different from the semigroup we seek equivariance to. In
particular this will be the case in
Section~\ref{sec:scale-translation-equivariance}, when we consider the
semigroup product of translations \emph{and} scalings. In theory the
issue is easily overcome, as changing the range of the sum from
$g\in G$ to $x\in X$ in \eqref{eq:semigroup-convolution} does not
change the equivariance property. In practice, that means defining a
semigroup action $R_u(f)(x)$. We propose to split this task into two
steps. First, we define the semigroup in which
\eqref{eq:semigroup-convolution} holds as it is, like in
Section~\ref{sec:scale-translation-equivariance}. Second, we introduce
a \emph{lifting} operator $\Lambda$ to map a function $f$ defined on $X$
into a function $\Lambda f$ defined on $G$, as it is done in Section~\ref{sec:scale-spaces}. The operator $H$ becomes then
\begin{equation*}
    \forall u\in G, \;\;\;  H(f)(u) = \sum_{g\in G} R_u(\Lambda f)(g) h(g).
\end{equation*}

Since $f$ and $H(f)$ now lie in different spaces, a more general
definition of equivariance is necessary: $H$ is equivariant to $G$ if
there exists two actions $R_u$ and $R^{\prime}_u$ such that
$H(R^{\prime}(f)) = R(H(f))$. Now we easily check that it is
sufficient for the lifting operator to be equivariant to $G$, that is,
$R_u \circ \Lambda = \Lambda \circ R^{\prime}_u$, to have equivariance of $H$. Indeed, in that case by omitting parentheses for readability,
\begin{equation}
    H (R^{\prime}_t f) (u) = \sum_{g\in G} R_u \Lambda R^{\prime}_t f  (g) h(g) = \sum_{g\in G} R_u R_t \Lambda f (g) h(g) = R_t(H f) (u).
\end{equation}
The advantage of this two-step approach is the richness of operators induced by the variety of possible liftings, as exposed in Section~\ref{sec:scale-spaces}.

\section{Scale and Translations Semigroup}
\label{sec:scale-translation-equivariance}

\subsection{Scale Cross-Correlation}
We focus on the semigroup product of scalings and translations $G = \mathcal{S} \rtimes \Z^2 = (\mathcal{S} \times \Z^2, \cdot)$, with $\mathcal{S} = \{\gamma^{-i} | i \in \mathbb{N} \}$, and $\gamma>1$ is an integer. The operation $\cdot$ is defined as $(s, x) \cdot (t, y) = (st, s^{-1} y + x)$. 
Assuming $\gamma=2$, the operator \eqref{eq:semigroup-convolution} applied to a signal $f$ at a point $(s, x)$ becomes \cite{worrall2019scale}
\begin{equation} \label{eq:scale_semigroup_corr}
    H(f)(2^{-k}, x) = (f \star_G h)(2^{-k},x) = \sum\limits_{l \geq 0} \sum\limits_{y \in \Z^2} f(2^{-k-l}, 2^k y + x) h(2^l, y).
\end{equation}


The operations here were defined for single channel images on $G$, but they can easily be applied to multichannel images. Let the input $f = (f_1, \dots, f_n) \in (\R^n)^G$ be a signal with $n$ channels. Assuming the output has $m$ channels, the filter is of the form $h: G \to \R^{n \times m}$. In this case \cite{worrall2019scale}, we compute the operator $H$ at channel $o \in \{1, \dots, m\}$ as $(f \star_G h)_o(2^{-k},x) \coloneqq \sum\limits_{c=1}^n (f_c \star_G h_{c, o})(2^{-k}, x)$.

Note that in \eqref{eq:scale_semigroup_corr}, because of the multiplicative constant $2^k$ in the spatial component, the receptive field (i.e. the region that the network ``sees'' of the input image at any given output position) of networks consisting of scale semigroup correlations are large. Indeed, a network consisting of $L$ scale cross-correlations with filters with dimensions $P \times K \times K$(i.e. the support of the filter is a grid $\{(p,k_1,k_2) | p \in \{2^0, 2^1, \dots, 2^{P-1}\}, 1 \leq k_1, k_2 \leq K \}$) has as a receptive field at each scale $s$ a square of sides $K + 2^{s+P-1} (L-1) (K-1)$, which is large when compared to a convolutional network with $L$ layers and $K \times K$ filters, which has a receptive field of size $L (K-1) + 1$. In other words, an architecture built on scale cross-correlations attains the same receptive field as a deep CNN with much smaller depth and number of parameters.

As anticipated in Section~\ref{sec:lifting}, we now have equivariant operators on functions $f:G \to \R^n$ and we need to apply this to images supported by a grid.
In the next sections we use the notion of scale-space to define lifting operators that map images to functions on the semigroup of scales and translations. With that we aim to define a neural network architecture that is equivariant with respect to re-scaling of 2D images.


%

\subsection{Scale-Spaces as Lifting Operators}
\label{sec:scale-spaces}

Following the definitions of \cite{heijmans02scale}, a family $\{ S(t):\mathcal{F}^{\R^2} \to \mathcal{F}^{\R^2} | t>0\}$ of operators on images is a \emph{scaling} if:
\begin{eqnarray}
    \label{eq:ss_def}
    S(1) = \mathrm{id},\;\;\;\;\;&
    \forall t, s > 0 \;\;S(t) S(s) = S(ts)
\end{eqnarray}
where $\mathrm{id}$ is the identity transform.
A scaling can be seen as an action of the group of continuous scalings, which is isomorphic to $(\R^+_*, \times)$.
An example is the family $S^{p, q}$, $p, q \geq 0$, given by
    $(S^{p, q}(t)f)(\bx) = t^q f\left( \frac{1}{t^p} \bx \right)$,
where $p$ and $q$ control rates of the spatial and contrast scaling, respectively. 

Let $S$ be a scaling and $\dot{+}$ a commutative operation such that $(\mathbb{R}^+_*, \dot{+})$ is a semigroup. Then a $(S, \dot{+})$ \emph{scale-space} is a family $\{ T(t) | t > 0\}$ of operators such that, for all $t, s > 0$ \cite{heijmans02scale}:
\begin{eqnarray}
    \label{eq:scale-space-def}
    T(t) T(s)  = T(t \dot{+} s), \;\;\; & \;\;\;
    T(t) S(t) = S(t) T(1). 
\end{eqnarray}
The property $T(t) S(s) = S(s) T(t/s)$, for all $t, s>0$, is a direct consequence of the second property~\cite{heijmans02scale}.
Here, in addition to  \eqref{eq:scale-space-def}, we assume that the scale-space $T(t)$ is translation-equivariant for all $t>0$ (i.e. $T(t) (L_z f) = L_z(T(t)f)$ where $L_z(f)(x) = f(x+z)$).
Thanks to the second property in \eqref{eq:scale-space-def}, a $(S^{p, 0}, \dot{+})$ scale-space $T$ defines an operator on images $f:\R^2 \to \R^C$
\begin{equation} \label{eq:lifting_scale}
    \forall (s, x) \in \mathcal{S} \times E \;\;\; (\Lambda f)(s,x) = (T(s^{-\frac{1}{p}})f)(x)
\end{equation}
such that, for all $(t,z) \in \mathcal{S} \times \Z^2$, $R_{(t, z)} \circ \Lambda = \Lambda \circ R_{(t, z)}'$,
where $R_{(s,x)}'$ is applied to an image on a continuous domain, $f:\R^2 \to \R^C$, as $(R_{(s,x)}'f)(y) = f(s^{-1} y + x)$. 
So, in order for $\Lambda$ to be our lifting operator we assume that the input $f$ is a function on a continuous domain.
In practice, we discretize $\Lambda$ and the input images.
With this, the morphological scale-spaces, as well as the Gaussian scale-space, being $(S^{\frac{1}{2}, 0}, \dot{+})$ scale-spaces, can be used as the lifting operators.


\subsubsection{Gaussian Scale-Space Lifting:}
The Gaussian scale-space is a $(S^{\frac{1}{2}}, +)$ scale-space defined by the family $T_\mathcal{G}(t)$. For all images $f \in \R^{\R^2}$ and points $x \in \R^2$, $T_\mathcal{G}$ can be computed as the convolution $(T_{\mathcal{G}}(t) f)(x) = (f * {\mathcal{G}}_t)(x)$ where $\mathcal{G}_t(x) = (2 \pi t)^{-1}\exp \left(- \frac{\lVert x \rVert^2}{2 t} \right)$.
This was the scale-space considered in \cite{worrall2019scale}. There, it was assumed that image $f$ has a maximum spatial frequency content. They model this by assuming that there exists a signal $f_0$ and a constant $s_0 > 0$ such that $f = (f_0 * \mathcal{G}_{s_0})$. 



\subsubsection{Quadratic Morphological Scale-Spaces:}
Morphological operators can form many different types of scale-spaces \cite{heijmans02scale}. In this paper we consider quadratic morphological scale-spaces.
The families of quadratic erosions and dilations by the structuring functions $q_t(x) = -\frac{\lVert x \lVert^2}{4 c t}$ $t>0$, given by
\begin{align}
    (T_{\varepsilon_q}(t)f)(x) & = \varepsilon_t(f)(x) = \inf\limits_{y \in \R^2} \left( f(x+y) - q_t(y)  \right), \; \text{ and,} \\
    (T_{\delta_q}(t)f)(x) & = \delta_t(f)(x) = \sup\limits_{y \in \R^2} \left( f(x-y) + q_t(y) \right), &
\end{align}
form $(S^{\frac{1}{2}, 0}, +)$ scale-spaces that can be regarded as morphological counterparts to the Gaussian scale-space \cite{boomgaard94}. Here, to increase flexibility, we consider a parameter $c>0$ learned by gradient descent with the rest of the parameters of the network.
It is also shown in \cite{heijmans02scale} that the openings $T_{\alpha_q}(t) = T_{\delta_q \circ \varepsilon_q} = \delta_t \circ \varepsilon_t$ and closings $T_{\beta_q}(t) = T_{\varepsilon_q \circ \delta_q} = \varepsilon_t \circ \delta_t$ by those structuring elements form $(S^{\frac{1}{2}, 0}, \vee )$ scale-spaces, where $\vee$ is the pairwise maximum, $a \vee b = \max \{a, b\}$.

\section{Experiments}
\label{sec:experiments}

\subsection{Image Classification}

The MNIST Large Scale dataset is built upon the MNIST dataset \cite{mnist} and was introduced to evaluate the ability of CNNs to generalize to scales not seen in the training set \cite{jansson2020unseenscales}. The dataset contains three training sets, tr1, tr2 and tr4, which consist of $50000$ samples from the MNIST dataset upscaled by factors one, two and four respectively. The remaining $10000$ samples from the original MNIST are used as validation sets, upscaled to match tr1, tr2 and tr4. In our experiments we use tr2 as the training set. The test set is re-scaled to the scales $2^{\frac{i}{4}}, i=-4,\dots,12$.

The scale-equivariant architecture used consists of the lifting layer, truncated at five scales, followed by $L=5$ scale cross-correlations layers\footnote{In our experiments, we use the implementation of the cross-correlation layer from \url{https://github.com/deworrall92/deep-scale-spaces}} and a global max-pooling, before a dense with a softmax activation.
The filters have dimension $1 \times 3 \times 3$, and each layer has $16, 16, 32, 32, 64$ feature maps and $1, 2, 1, 2, 1$ strides, respectively, using Batch Normalization and ReLU activations. The architecture is similar to the one used in \cite{jansson2020unseenscales}.
By taking a max-pooling over all scales and spatial positions we attempt to make the model invariant to the action of the semigroup $G$. This makes so that the output is at least as high as the output of the same model after the action of the $G$. So if at a certain scale the features are indicative of a certain class, the output should also be indicative of that class.

We compare liftings that use scale-spaces: the Gaussian scale-space $T_\mathcal{G}$, the quadratic dilation and closing scale-spaces\footnote{We do not use erosion and opening scale-spaces because the bright objects in a dark background would be erased by the anti-extensive operators. For a dataset without a well-defined polarity, self-dual operators could be used instead.} $T_{\delta_q}$ and $T_{\beta_q}$ and the scale-space $T_{\text{id}}(t) = \text{id}$ for all $t>0$. We also compare with a CNN with size similar to the equivariant models'. 
For this proposal, one can measure the Euclidean distance between the features obtained from the same images at different scales to quantify the quality of invariance of the model. The distance is normalized by the norm of the inputs. For the purposes of this experiment, we consider the features of the whole dataset as a single vector.

\begin{figure}
    \centering
    \begin{subfigure}{.49\textwidth}
        \includegraphics[width=\textwidth]{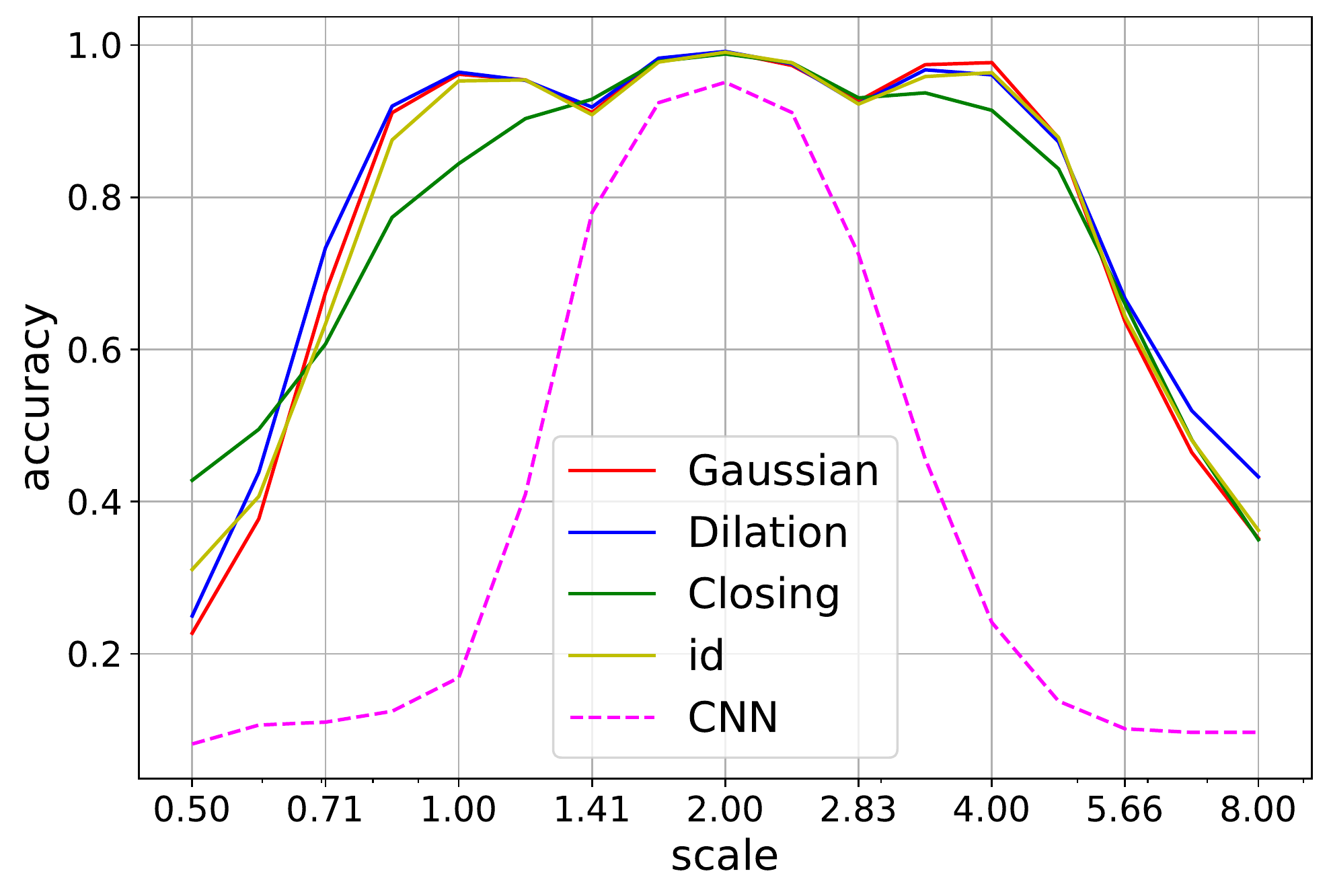}
        \caption{Accuracies of the models at different scales.}
    \end{subfigure}
    \begin{subfigure}{.49\textwidth}
        \includegraphics[width=\textwidth]{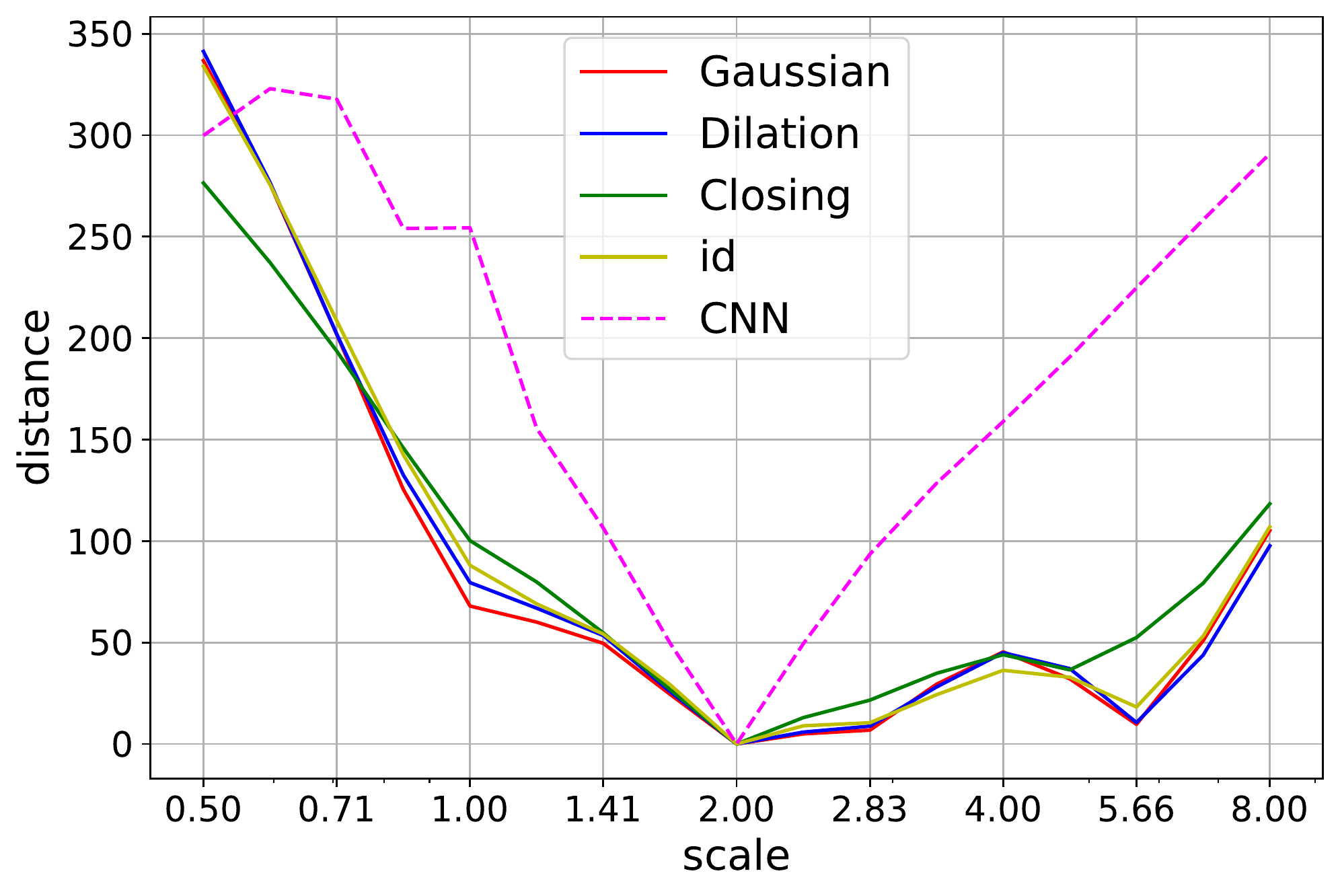} 
        \caption{Distances between the features in the training scale and all the test scales.}
    \end{subfigure}
    \centering
    \caption{Results on MNIST Large Scale experiment averaged from five initializations of the models.}
    \label{fig:invariance_mnist}
\end{figure}

Fig. \ref{fig:invariance_mnist}(a) shows the accuracies of the models when tested with different scales and in Fig. \ref{fig:invariance_mnist}(b) the distances between the features obtained from the same input image at different scales.
As expected the equivariant models outperform the CNN model in terms of generalization and even at the training scale. The difference between the peaks of the equivariant models and the CNN models may be related to the difference in the receptive field. In this experiment, the equivariant models performed similarly, with performance peaks at scales one, two and four which are one scaling upwards or downwards from one another. The distances in Fig. \ref{fig:invariance_mnist}(b) are mostly consistent with the accuracies and smaller in the equivariant models.

\subsection{Image segmentation}
\label{sec:seg}

In this section we perform an experiment on image segmentation where the objects in the image are re-scaled independently of one another. Unlike in the classification problem, it would be difficult to obtain the images from the dataset by means of data augmentation. In this problem the network benefits from being locally invariant to re-scaling\footnote{We say that an operator $\psi$ on images is locally invariant w.r.t. to $R^\prime_{(s,x)}$ if $\forall (s,x) \in G$ $\psi R^\prime_{(s,x)}f = R^\prime_{(1,s^{-1}x)} \psi f$.}. We apply a max-pooling over the scale dimension, i.e. the operator $M(f)(x) = \max\limits \{ f(s,x) | s \in \{2^0, 2^1, 2^2, \dots, 2^N\} \}$. This means that the activations used by the softmax layer are the highest for each scale, and should in practice make a locally invariant-model.

\begin{figure}
    \centering
    \begin{subfigure}{.2\textwidth}
        \includegraphics[width=\textwidth]{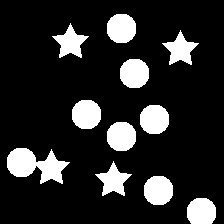}
        \caption{}
    \end{subfigure}
    ~
    \begin{subfigure}{.2\textwidth}
        \includegraphics[width=\textwidth]{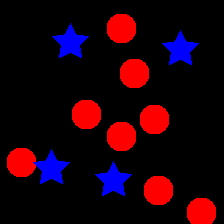}
        \caption{}
    \end{subfigure}
    ~
    \begin{subfigure}{.2\textwidth}
        \includegraphics[width=\textwidth]{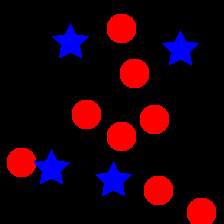}
        \caption{}
    \end{subfigure}
    
    \begin{subfigure}{.2\textwidth}
        \includegraphics[width=\textwidth]{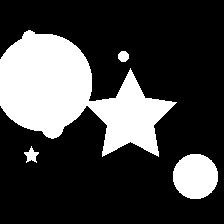}
        \caption{}
    \end{subfigure}
   ~   
    \begin{subfigure}{.2\textwidth}
        \includegraphics[width=\textwidth]{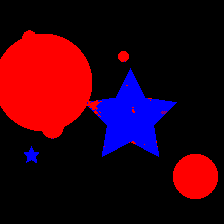}
        \caption{}
    \end{subfigure}
    ~
    \begin{subfigure}{.2\textwidth}
        \includegraphics[width=\textwidth]{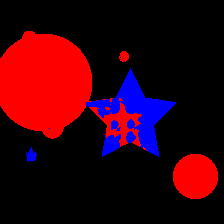}
        \caption{}
    \end{subfigure}
    \caption{Example images from the (a) training and (d) test sets of the segmentation experiments, and segmentation results using equivariant models with the (b)(e) proposed dilation and (c)(f) Gaussian scale-spaces. Pixels in red are classified as \emph{disk}, those in blue as \emph{star}.}
    \label{fig:seg_examples1}
\end{figure}

The dataset we use in this experiment consists of $224 \times 224$ synthetic binary images of shapes such as Fig. \ref{fig:seg_examples1} which are divided in three classes: disks, stars and the background. In the training set, only one scale is present, like in Fig. \ref{fig:seg_examples1}(a). We construct test sets where each shape is re-scaled by a factor uniformly sampled from the interval $[2^{-i}, 2^{i}]$, in which we use $i = 1,2$. Fig. \ref{fig:seg_examples1}(b) shows an example of a test image.
The train set contains 10000 images and the test sets contain 500 images each. The experiment is repeated ten times, each time generating a different training/test set pair.

The architecture chosen for the equivariant models consists simply of six layers of semigroup cross-correlations. Because the scale cross-correlation has a naturally large receptive field, subsampling is not necessary. The output of the network is a three-channel image with the scores for each class, and the class is chosen as the coordinate with the greatest score.
To quantitatively evaluate the models, the Intersection over Union(IoU), or Jaccard index, between the ground truth image and the predictions is used.
As baselines, we compare the models to a CNN with the same number of layers and a similar size and number of parameters, and also to a U-Net \cite{ronneberger2015unet} architecture.

In Table \ref{tab:jacard} we compare the IoU obtained from different models. We see that CNN performs badly, compared to the equivariant models, even in the training set scale. This is partially attributed to the fact that the receptive field of the CNN is not as large, although having the same number of layers and a similar number of parameters. As expected, the equivariant models outperformed the CNN architectures and the U-Net architecture in the generalization to other scales.

\begin{table}
    \centering
    \scalebox{.8}{
    \begin{tabular}{|c|c|c|c|c|c|c|}
        \hline
        Scales              & Gaussian Lifting      & Dilation Lifting      & Closing Lifting       & Id Lifting           & U-Net               & CNN \\
        \hline                                                                                                              
        $1.$                & $0.9929 \pm 0.0006$   & $0.9929 \pm 0.0006$   & $0.9929 \pm 0.0005$   & $0.9927 \pm 0.0008$  & $0.9927 \pm 0.0006$ & $0.9083 \pm 0.0006$ \\
        $[\frac{1}{2}, 2]$  & $0.92 \pm 0.06$       & $0.97 \pm 0.01$       & $0.89 \pm 0.06$       & $0.91 \pm 0.03$      & $0.86 \pm 0.02$     & $0.68 \pm 0.01$   \\
        $[\frac{1}{4}, 4]$  & $0.88 \pm 0.07$       & $0.93  \pm 0.02$      & $0.86 \pm 0.05$       & $0.88 \pm 0.03$      & $0.70 \pm 0.04$     & $0.627 \pm 0.008$ \\
        \hline
    \end{tabular}
    }
    \caption{IoU between the ground truth images and the predictions obtained from equivariant models with different liftings, trained on images where objects only appear at scale one.}
    \label{tab:jacard}
\end{table}

\begin{figure}
    \centering
    \begin{subfigure}{\textwidth}
        \centering
        \includegraphics[width=.18\textwidth]{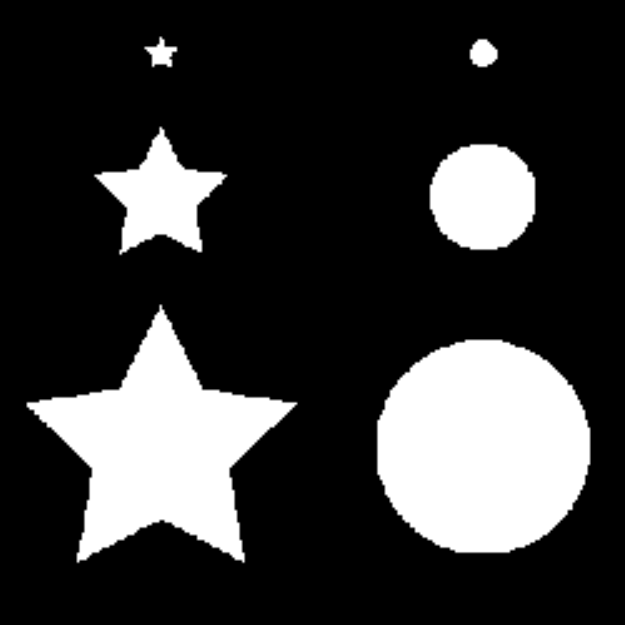}
        \includegraphics[width=.18\textwidth]{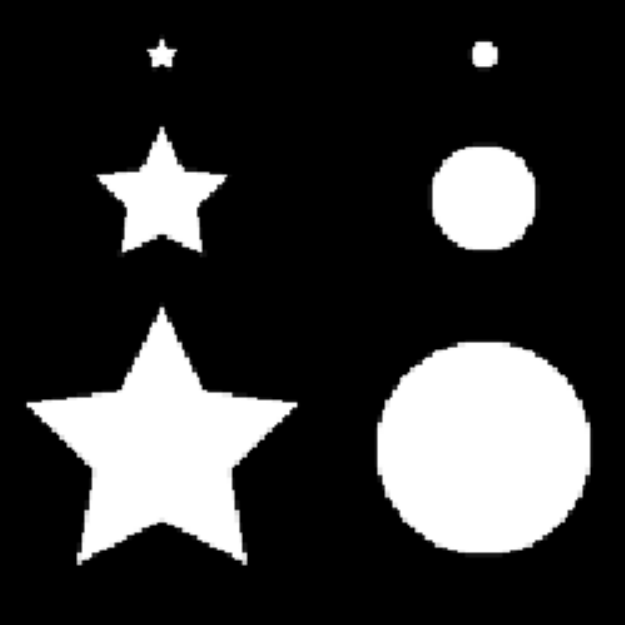}
        \includegraphics[width=.18\textwidth]{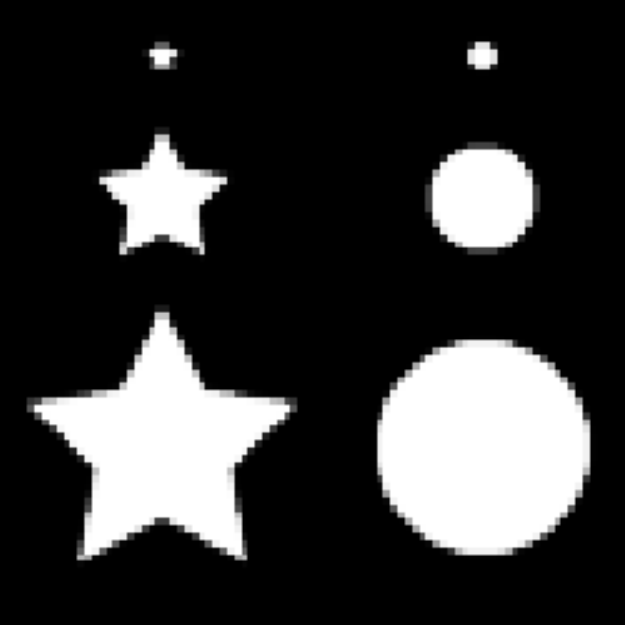}
        \includegraphics[width=.18\textwidth]{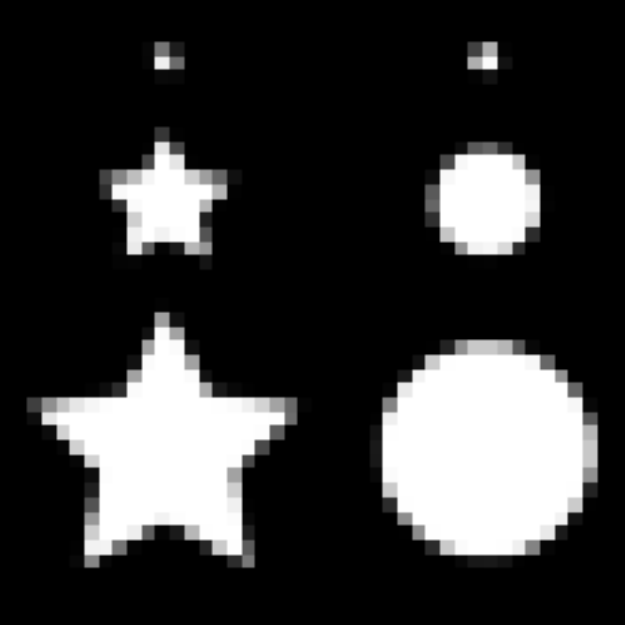}
        \includegraphics[width=.18\textwidth]{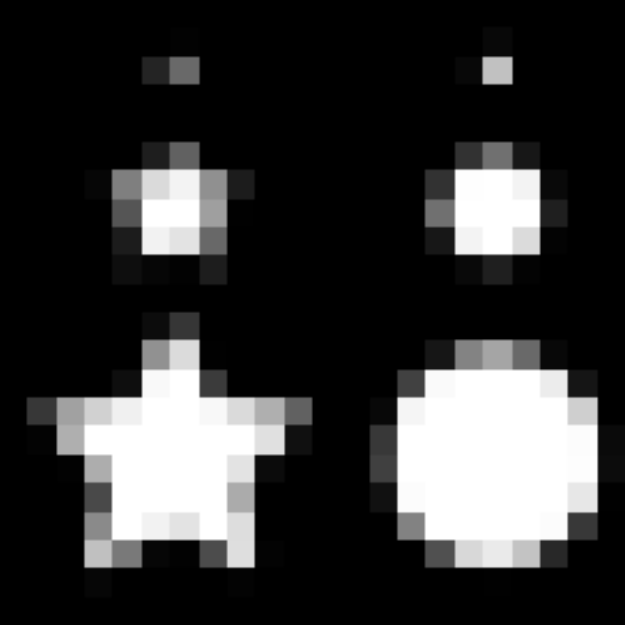}
        \caption{Gaussian scale-space}
    \end{subfigure}%
    
    \begin{subfigure}{\textwidth}
        \centering
        \includegraphics[width=.18\textwidth]{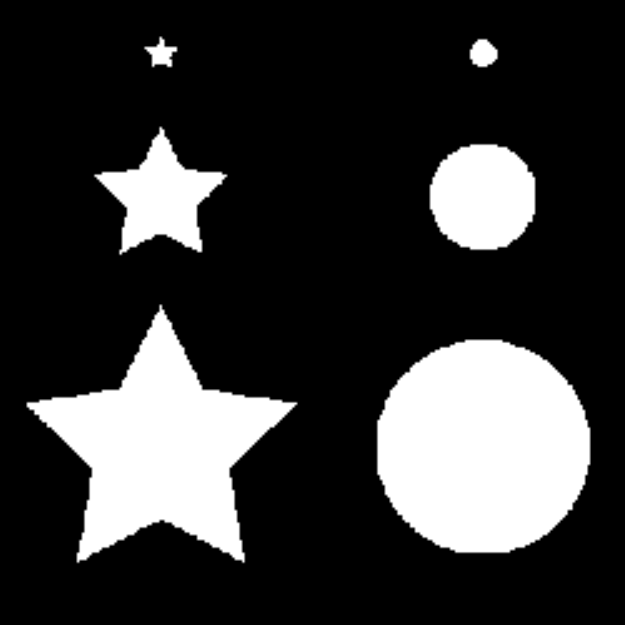}
        \includegraphics[width=.18\textwidth]{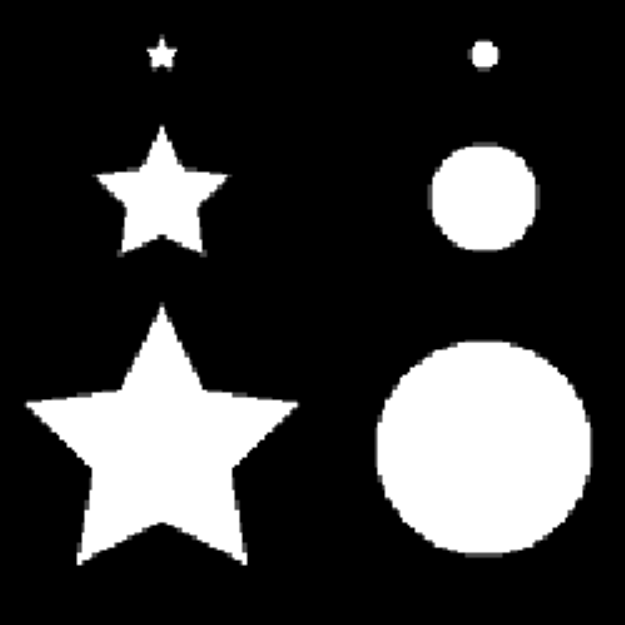}
        \includegraphics[width=.18\textwidth]{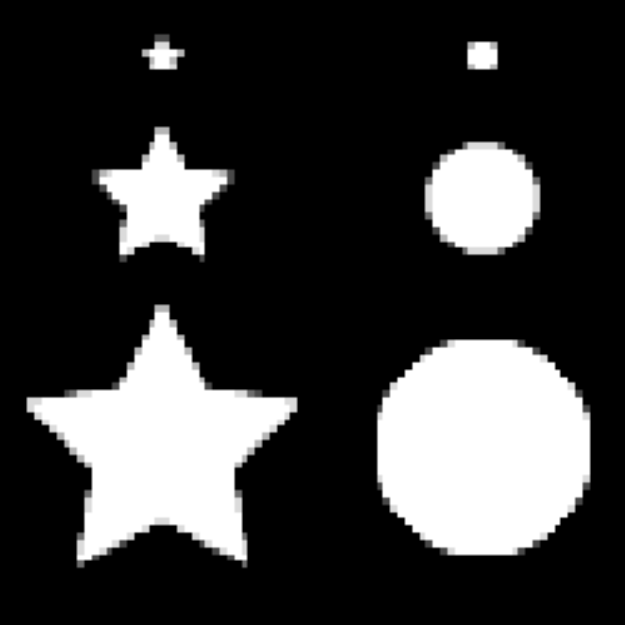}
        \includegraphics[width=.18\textwidth]{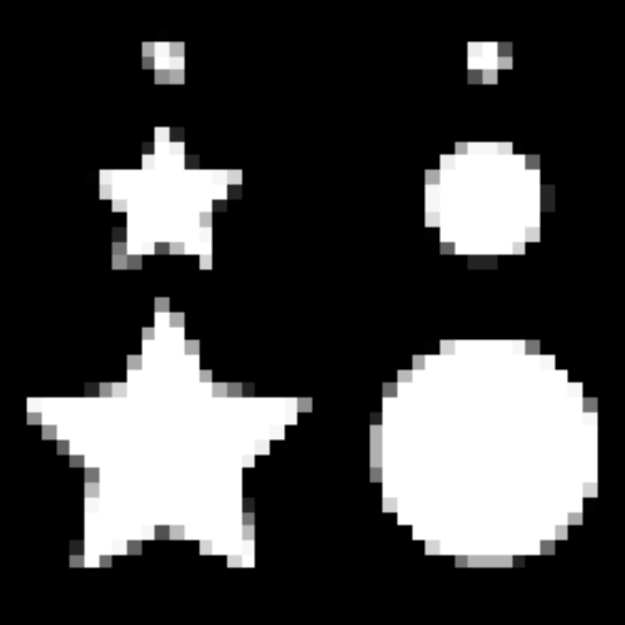}
        \includegraphics[width=.18\textwidth]{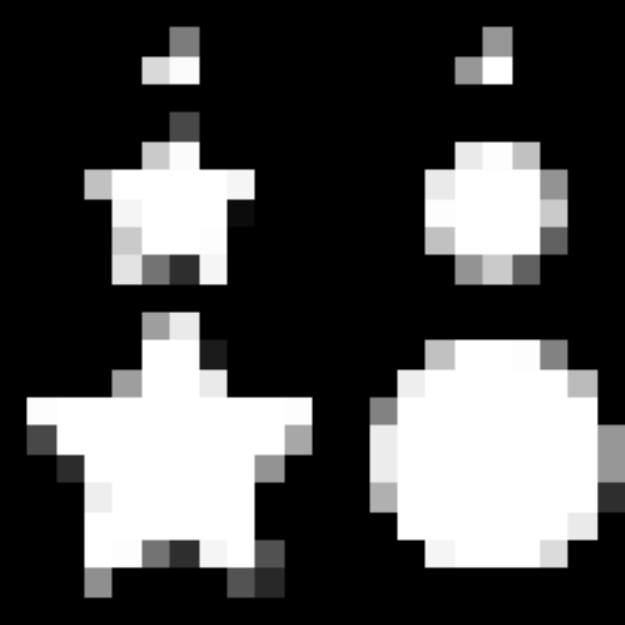}
        \caption{Quadratic Dilation scale-space}
    \end{subfigure}%
    \caption{The same image after being processed by the Gaussian (a) and quadratic dilations (b) scale-spaces and being subsampled by factors $2^i$ $i=0,1,2,3,4$.}
    \label{fig:seg_examples2}
\end{figure}

To analyse why the dilation is suited to this particular dataset, we can analyse the effect of applying a discrete re-scaling, i.e. a subsampling to the objects processed by the scale-spaces. 
In Fig. \ref{fig:seg_examples2} we see the difference between a Gaussian and dilation lifting followed by a subsampling operator. Indeed, the persistence of concavities of the star shapes makes it easier to distinguish the objects in the last images.

\section{Conclusions and Future Work}
In this paper we presented a generalization of the scale-equivariant models of Worrall and Welling \cite{worrall2019scale} based on the general definition of scale-spaces given in \cite{heijmans02scale}. The models obtained from this approach with different scale-spaces are evaluated in experiments designed to test invariance to change in scales. In our experiments, the generalization to new scales of the models based on scale-spaces surpassed the CNN baselines, including a U-Net model.
We see that changing the type of scale-space in the architecture of \cite{worrall2019scale} can induce a change in the performance of the model. In the datasets used, where the geometric information of the images is important, the dilation scale-space model compared very favorably to the Gaussian one.
Regarding future works, we note that only the second property of the scale-space definition \eqref{eq:scale-space-def} is necessary in order for operators to be equivariant, which means that some operators that fulfill it but are not scale-spaces, such as the top-hat transform, can be used as lifting operators.
It would also be interesting to compare these models with the invariance obtained from data augmentation and possibly combine the two approaches in a complementary way.
Additionally, future works will explore the use of the proposed morphological scale-spaces on other types of data like 3D point clouds \cite{asplund2019mathematical}, graphs \cite{Blusseau18}, and high dimensional images.

\bibliographystyle{splncs03}
\bibliography{references}
\end{document}